\documentclass[twocolumn,preprintnumbers,amssymb,floatfix,aps,prl]{revtex4}
\usepackage{graphicx}% Include figure files
\usepackage{dcolumn}% Align table columns on decimal point
\usepackage{bm}% bold math
\begin{document}

\title{Microscopic Model for High-spin vs. Low-spin ground state in 
$[Ni_2{M(CN)_8]}$ ($M=Mo^V,~W^V,~Nb^{IV}$) magnetic clusters}

\author{Rajamani Raghunathan$^\S$, Jean-Pascal Sutter$^\#$, 
Laurent Ducasse$^\ddagger$, C\'edric Desplanches $^\dagger$ and S. Ramasesha$^\S$
\footnote{E-mail: rajamani@sscu.iisc.ernet.in; sutter@lcc-toulouse.fr; 
l.ducasse@lpcm.u-bordeaux1.fr; desplan@icmcb-bordeaux.cnrs.fr; 
ramasesh@sscu.iisc.ernet.in}}

\affiliation {$^\S$Solid State and Structural Chemistry Unit, Indian Institute of Science, Bangalore 560 012, India.\\
$^\#$Laboratorie de chimie de coordination du CNRS, Universit\'e Paul Sabatier, 
205, Route de Narbonne, 31077 Toulouse, France.\\
$^\ddagger$ Laboratoire de Physico-Chimie Mol\'eculaire, UMR 5803 du CNRS, Universit\'e Bordeaux I, 351 Cours de la Lib\'eration, 33405 Talence, France.\\
$^\dagger$Institut de Chimie de la Mati\'ere Condens\'ee de Bordeaux, CNRS, 87 Avenue Dr. Schweitzer, F-33608 PESSAC, France.}

\begin{abstract}
Conventional superexchange rules predict ferromagnetic exchange interaction 
between $Ni(II)$ and $M$ ($M=Mo^V,~W^V,~Nb^{IV}$). Recent experiments 
show that in some systems this superexchange is antiferromagnetic. To 
understand this feature, in this paper we develop a microscopic model for 
$Ni(II)-M$ systems and solve it exactly using a valence bond approach. 
We identify the direct exchange coupling, the splitting of the 
magnetic orbitals and the inter-orbital electron repulsions, on the 
$M$ site as the parameters which control the ground state spin of various 
clusters of the $Ni(II)-M$ system. We present quantum phase diagrams 
which delineate the high-spin and low-spin ground states in the parameter 
space. We fit the spin gap to a spin Hamiltonian and extract the effective 
exchange constant within the experimentally observed range, for 
reasonable parameter values. We also find a region in the parameter space 
where an intermediate spin state is the ground state. These results indicate 
that the spin spectrum of the microscopic model cannot be reproduced by a 
simple Heisenberg exchange Hamiltonian. 

\centerline{(Received date : \today)}
\end{abstract}

\maketitle

\section{I. Introduction}

\begin{figure}[ht]
\begin{center}
\includegraphics[width=2.5in, height=2.0in]{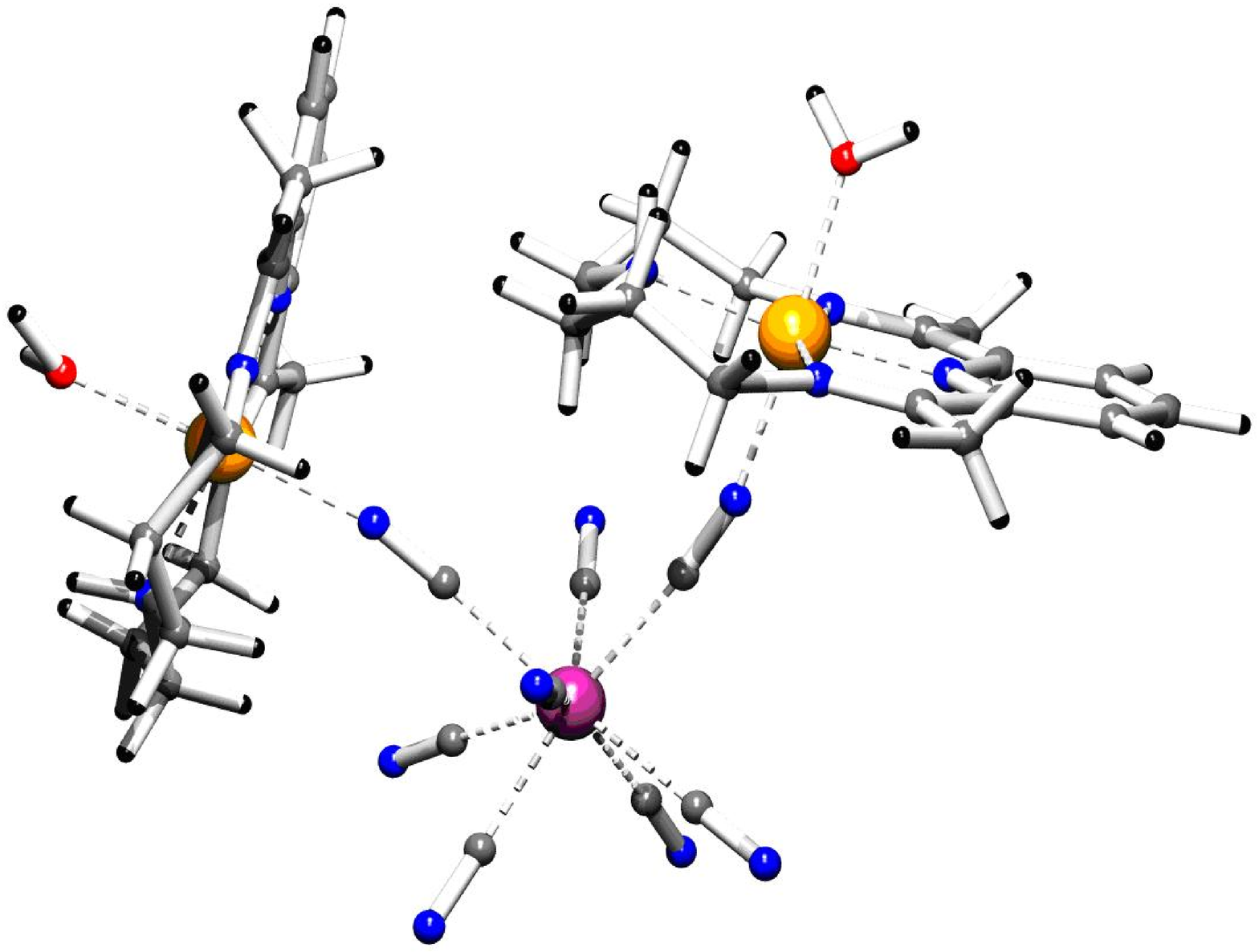}
\hspace{1.0cm}\includegraphics[width=2.5in, height=2.0in]{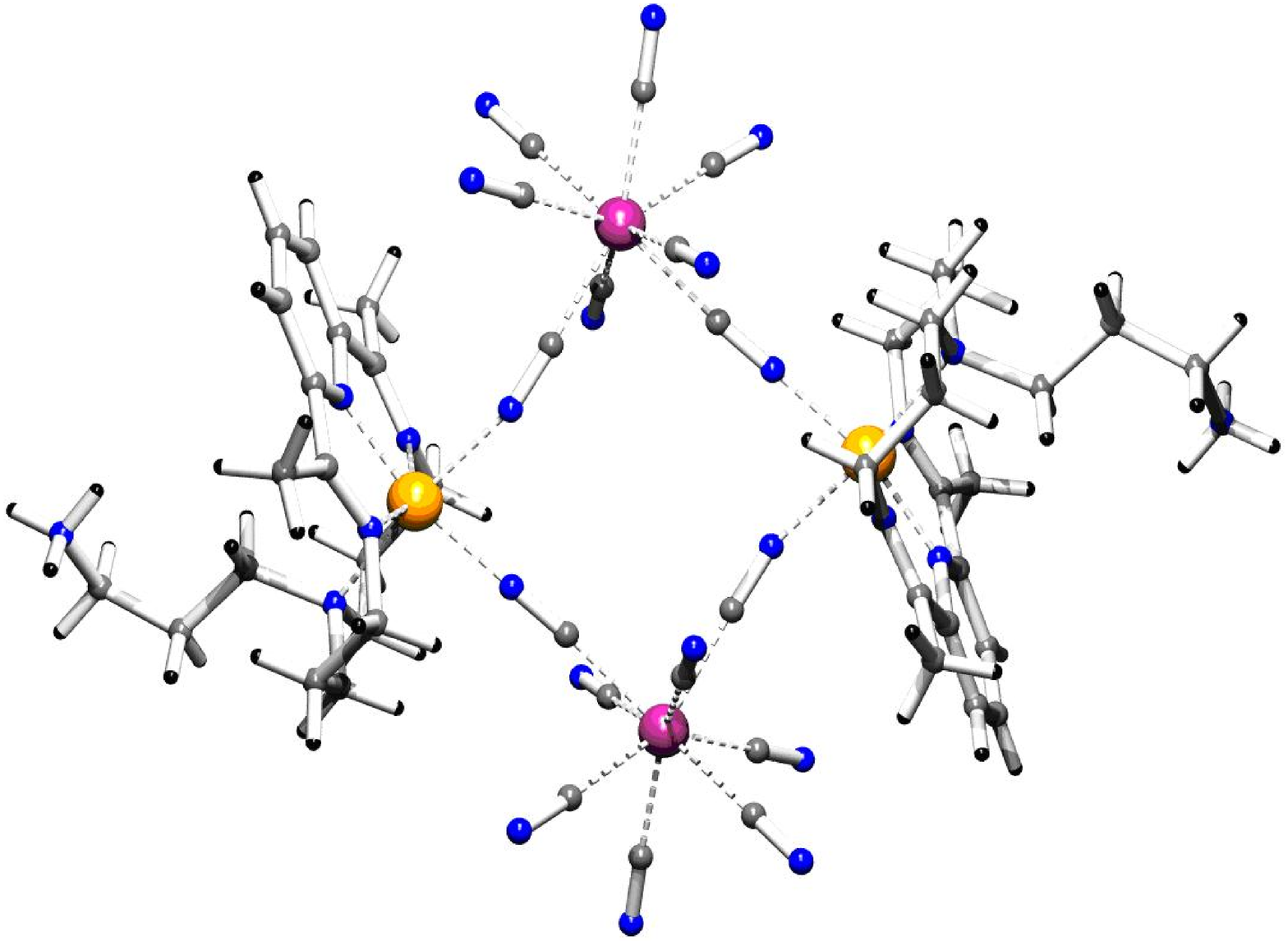}
\caption[]{Views of the supramolecular compounds top:  
$[\{NiL\}_2\{M(CN)_8\}]^+$ and bottom : $[\{NiL'\}_2\{M(CN)_8\}_2]$.}
{\label{supramol}}
\end{center}
\end{figure}

In the expanding field of molecular magnetism, transition metal complexes 
have been the main focus of study. Room temperature bulk magnetization has 
been achieved with a hybrid {organic radical/V(II)} framework \cite{manriquez}
and also with the {Cr(III)/V(II)} Prussian blue type compound \cite{ferlay,
holmes}. Besides, interesting conjectures such as the Haldane conjecture 
\cite{haldane} have been verified in molecular systems\cite{verdauger}. 
The interest in 
molecular magnetism has expanded in recent times to include low-dimensional 
architectures exhibiting slow relaxation of magnetization known as Single 
Molecule Magnets (SMMs) \cite{gatteschi}, conducting magnets \cite{coronado} 
and light-triggered magnets \cite{sato}. The large majority of these 
molecular compounds are based on paramagnetic 3d metal ions, the second 
and third row transition metal ions have been envisaged only recently 
for the construction of magnetic supramolecular compounds \cite{corine-VCH}. 
These ions appear however very promising in molecular magnetism yielding 
magnets with fairly high ordering temperatures \cite{tanase}, new SMM's 
\cite{song} and efficient systems with light triggered magnetization changes 
\cite{corine}. The 4d and 5d ions are characterized by spatially more 
extended valence orbitals, the extension following the trend $3d~<~4d~<~5d$. 
A consequence of this ordering of the d-orbital is that the on-site electron 
repulsion is decreased as we go down the column in the periodic table. Besides, 
the metal-ligand bonds become more covalent resulting in more efficient electron 
delocalization. A series of experimental results obtained on simple 
cyano-bridged heterometallic species formed by self-assembling of an 
octacyanometallate ($\{Mo^V(CN)_8\}^{3-}$ or $\{W^V(CN)_8\}^{3-}$) with a 
$\{Ni^{II}L\}$ complex (L = macrocyclic ligand) showed that the nature of the 
effective superexchange through the cyano-bridge depends upon the actual spin 
topology of the complex that is being studied. For instance, whereas 
significant ferromagnetic interactions are found for linear \{$Ni-M-Ni$\} 
compounds ($M$ = $Mo^V$ or $W^V$ in the [\{$M(CN)_8$\}] unit), 
an antiferromagnetic behavior is observed for a cyclic tetranuclear 
compound (Figure \ref{supramol}) \cite{sutter-unp}. A related larger spin 
clusters of formula $[\{NiL\}_{12}\{Nb(CN)_8\}_6]$ exhibits an even more 
complex magnetic behavior \cite{NbNi}. 

\begin{figure*}[t]
\includegraphics[height=1.7in,width=5.0in]{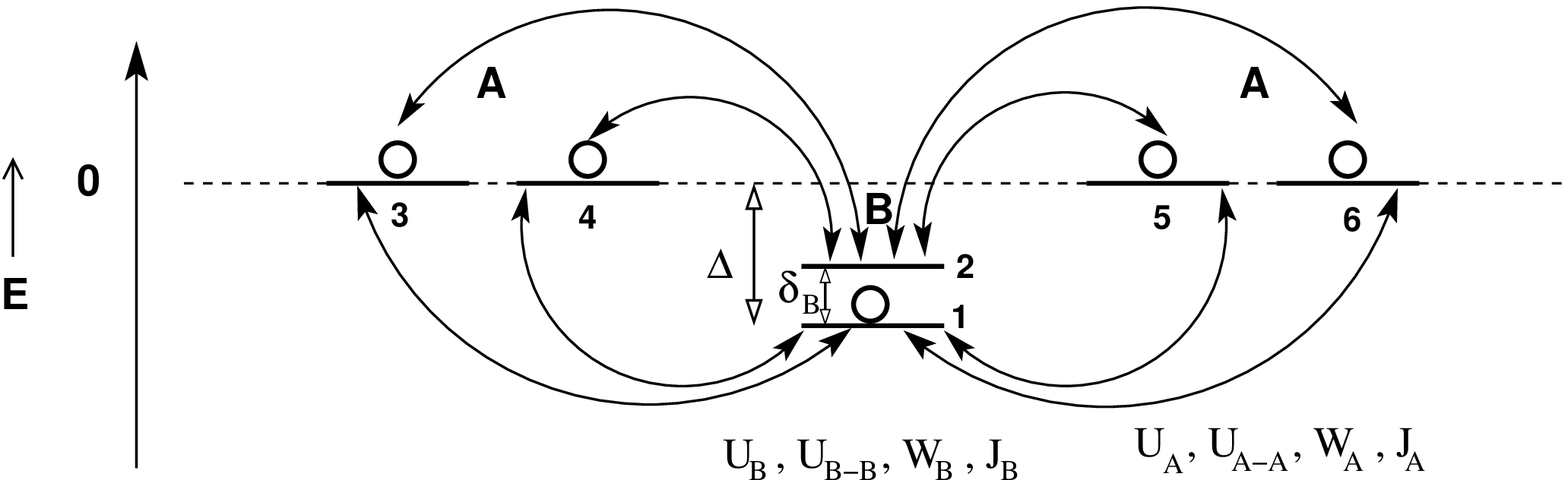}
\caption{Schematic diagram of the active orbitals and the interactions
in the A-B-A system. The transfer terms included in the model are
shown by filled double-headed arrows between pairs of orbitals. Since, 
only two orbitals are considered on each site, the interaction terms are 
labeled by the site indices (A/B).}
{\label{fig-model} }
\end{figure*}

In this paper, we explain this puzzling nature of the superexchange 
interaction between two transition metal ions $A$ and $B$, each containing 
two nearly degenerate orbitals, by developing a microscopic model which 
naturally admits antiferromagnetic exchange interactions between them. The ion 
$A$ has two electrons while $B$ has one electron; these correspond to the 
case of $Ni^{II}$ for the $A$ ion and either $Mo^V$, $W^V$ or $Nb^{IV}$ 
for the $B$ ion. 
The parameters entering the definition of the microscopic model are the 
on-site and inter-site parameters. The on-site parameters are both one and 
two-electron parameters. The one-electron parameters are energy differences 
between the active magnetic orbitals on $A$ and $B$, and the splitting of 
the two orbitals on $B$ \cite{footnote}. The two electron parameters are 
the electrostatic on-site Hubbard repulsion terms and the inter-orbital 
Coulomb and exchange integrals. The inter-site one-electron parameters 
are the transfer integrals and the inter-site two-electron Coulomb 
interactions in the zero differential overlap approximation.  
We identify the critical interaction parameters that control the sign of the 
superexchange interaction and present a quantum phase diagram that shows a 
crossover in the ground state from effective ferromagnetic interaction to 
effective antiferromagnetic interaction between the ions. We show this 
behavior in different organizations of the $AB$ system such as $A-B$, linear 
$A-B-A$, $B-A-B$, $A-B-A-B$, and cyclic $A-B-A-B$ clusters. In the next 
section we introduce the microscopic model for superexchange 
in the $Ni^{II}_{S=1}-M^{V}_{S=1/2}$ system. This is followed by a section 
in which we describe the valence bond (VB) method for solving the model. 
The last section deals with the results and their discussion.

\section{II. Model Hamiltonian for Superexchange in AB Systems}

The nature of exchange interaction between two ions is the outcome of a
competition between delocalization of the electrons which reduces the
kinetic energy and direct exchange interactions which reduces the 
electron-electron repulsion energy for parallel spin alignment. Usually, 
the kinetic energy is decreased if the electrons are in a low-spin state; 
in the high spin state, the delocalization of electrons is blocked because 
of the Pauli exclusion. Besides, low-spin configuration also usually affords 
more phase space for electron delocalization. This is why the kinetic 
exchange is usually antiferromagnetic in nature. However, when there 
is a near degeneracy of the partially occupied orbitals on an ion, then 
delocalization is possible even when the spins are aligned parallel and 
indeed, the direct exchange interaction favours parallel electron spin 
alignment on such degenerate site. In this case, the final outcome cannot 
be easily predicted and depends upon the actual values of the interaction 
parameters. Thus, a microscopic model for explaining observed sign of the
exchange interaction should include direct exchange interactions between
nearly degenerate orbitals on a given site and the relative splitting of the
degenerate orbitals.  Besides these terms, another interaction term of 
considerable import is the strength of the intra-orbital electron 
repulsions relative to the strength of on-site inter-orbital electron 
repulsions. This is because,
if the inter-orbital intra-site electron repulsions is weak compared 
to the intra-orbital electron repulsion, it will favour single occupancy
of two orbitals on the same site over double occupancy of one of the orbitals.
Occupancy of the orbitals controls the nature of the superexchange process,
thus the relative strengths of on-site inter and intra orbital repulsions
become very important. The other interaction terms which are comparable
in strength to these repulsion integrals are the intersite electron-electron 
interactions within the zero differential overlap approximation and the
intra-site electron repulsion between a charge density in one orbital and
a charge density in the overlap cloud of two orbitals, which we call the 
$W$ term. The model Hamiltonian for investigating the 
superexchange interaction can be written as:

\begin{eqnarray}
\hat{H} & = & \sum_i \epsilon_i \hat n_i + \sum_{<ij>} t_{ij}(\hat E_{ij} 
 + \hat E_{ji}) \nonumber \\
&& + \sum_i {\frac {U_i}{2}}\hat n_i (\hat n_i -1) + \nonumber\\
&& \sum_{i,i^\prime}
\{ U_{i,i^\prime} \hat n_i \hat n_{i^\prime} + {\frac{W_{i,i^\prime}}{2}}
\left[(E_{i,i^\prime}+E_{i^\prime,i})(\hat n_i+ \hat n_{i^\prime})\right. \nonumber\\
&&\left. +(\hat n_i+ \hat n_{i^\prime})(E_{i,i^\prime}+E_{i^\prime,i})-
2((E_{i,i^\prime}+E_{i^\prime,i})\right] \nonumber \\
&&+{\frac {J_{i,i^\prime}}{2}}\left(E_{i,i^\prime}E_{i,i^\prime}+
E_{i^\prime,i}E_{i^\prime,i}+E_{i,i^\prime}E_{i^\prime,i}\right. \nonumber\\
&&\left. +E_{i^\prime,i}E_{i,i^\prime}-\hat n_i -\hat n_{i^\prime}\right)\} 
+ \sum_{<ij>}{\frac {V_{ij}}{2}} \hat n_i \hat n_j; \nonumber\\ 
&&\hat n_i = \sum_\sigma a^\dagger_{i,\sigma} a^{}_{i,\sigma};~
\hat E_{i,j} = \sum_\sigma a^\dagger_{i,\sigma} a^{}_{j,\sigma}
\label{H_Superexchange}
\end{eqnarray}

\noindent where, the operators $a^\dagger_{i,\sigma}$ ($a_{i,\sigma}$) creates
(annihilates) an electron in orbital $i$ with spin $\sigma$.
The first line corresponds to the non-interacting part of the 
Hamiltonian with $\epsilon_i$ being the energy of the $i^{th}$ orbital,
and $t$ the transfer integral between an orbital on one site
and another on the neighbouring site. All values of $t$ are assumed to
be the same and the orbital energy of the A type atoms are fixed at zero,
while the orbital energies of the B type atoms are $-\Delta$ and ($-\Delta
+\delta_B$) as shown in Fig. \ref{fig-model}. The second line corresponds 
to the intra-orbital interaction
term, with $U_i$ being the Hubbard parameter. The remaining lines except the last 
represent the inter-orbital on-site electron repulsion terms, $U_{i,i^\prime}
= [ii|i^\prime i^\prime ] = \int \int \phi_i^*(1) \phi^{}_i(1) 
\frac{1}{r_{12}} \phi_{i^\prime}^*(2) \phi^{}_{i^\prime}(2) d^3r_1 d^3r_2$,
$W_{ii^\prime} = [ii|ii^\prime] = \int \int \phi_i^*(1) \phi^{}_i(1) 
\frac{1}{r_{12}}\phi_i^*(2) \phi^{}_{i^\prime}(2) d^3r_1 d^3r_2$ and 
$J_{ii^\prime} = [ii^\prime|ii^\prime] = \int \int \phi_i^*(1) \phi^{}
_{i^\prime}(1) \frac{1}{r_{12}} $ $\phi_i^*(2) \phi^{}_{i^\prime}(2) 
d^3r_1 d^3r_2$, where $J_{ii^\prime}$ is the exchange integral. 
Since we deal with only two types of ions $A$ and $B$, and only two orbitals
on each ion, we label these as $U_A$, $U_{A-A}$, $W_A$ and $J_A$ for the
$A$ ions and similarly for the $B$ ions. The 
last line corresponds to the inter-site inter-orbital repulsion integral
$V_{ij} = [ii|jj] =  \int \int \phi_i^*(1) \phi^{} _{i}(1) \frac{1}{r_{12}} 
\phi_j^*(2) \phi^{}_{j}(2) d^3r_1 d^3r_2$ and is parametrized using 
Ohno parametrization \cite{ohno} and appropriate scaling \cite{sumit}. 

\section{III. Solution of the Superexchange Model}

\begin{figure}
\centerline{\includegraphics[height=4.3in,width=3.4in]{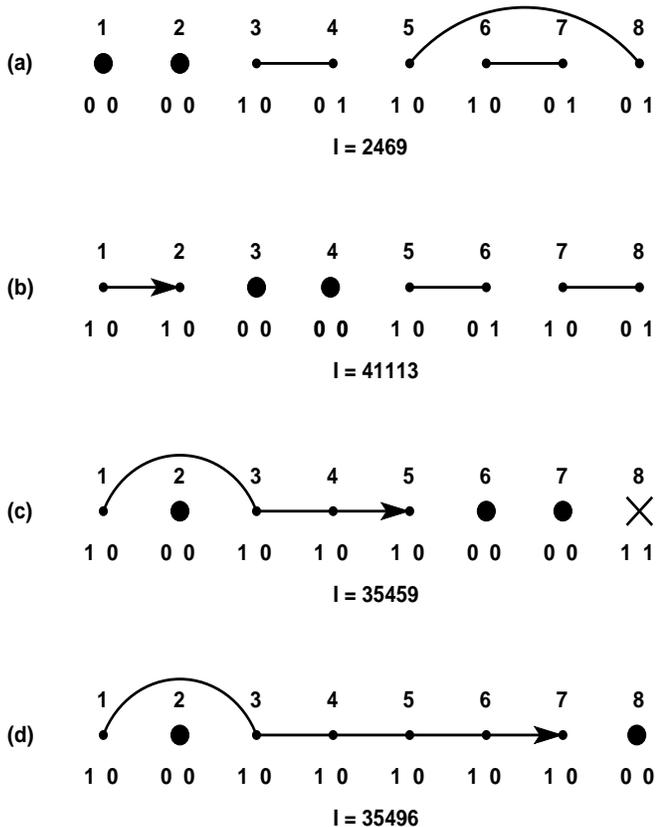}}
\caption[]{\label{figvb} Representative VB diagrams in (a) S=0, (b) S=1,
(c) S=2 and
(d) S=3 subspaces. Dots ($\bullet$)
represent empty sites and crosses ($\times$)
represent doubly occupied sites. In each VB diagram, numbers 1 through 8
are the site indices and the values at the bottom against each site gives
the bit state of the bits representing the site in the integer I.
The bit-pattern and the value of the integer I, representing the VB diagram
are also shown.}
\end{figure}

The Hamiltonian in eqn. \ref{H_Superexchange} is nonrelativistic and 
hence conserves total spin. Since we are interested in the solution 
of the model in different total spin sectors, we use the valence bond 
technique for solving the model. In this technique, the complete and 
linearly independent basis states 
with a given total spin can be generated using explicit spin pairings 
\cite{SR-VB}.  For example, if singly occupied sites $i$ and $j$ are 
spin paired, then a line is drawn between sites $i$ and $j$ in the VB 
diagram to indicate the spin pairing, $(\alpha_i\beta_j - \beta_i
\alpha_j)/ \sqrt 2, ~i<j$. We say that the line begins at site $i$ and 
ends at site $j$. If the spins at sites $i_1, i_2 \cdots i_l$ are not 
paired, we pass an arrow through these sites in the VB diagram and this 
denotes the state $\alpha_{i_1} \alpha_{i_2} \cdots \alpha_{i_l}$. Because 
the Hamiltonian conserves $S^z_{total}$ besides $S^2$, it is sufficient 
to work in the subspace $M_S = S$.  In the $S=0$ subspace, a VB diagram has
either empty and doubly occupied sites (represented by dots and crosses 
respectively) or lines between singlet paired singly occupied sites.
A VB diagram involving $N$ orbitals can be drawn by arranging the orbitals 
at the vertices of a regular $N$-gon and drawing straight lines between 
vertices, the electrons at whose orbitals are singlet paired. The 
Rumer-Pauling rule states that all such VB diagrams with no intersecting
lines form a complete and linearly independent set \cite{rumer}. The 
complete and linearly independent set of VB states in nonzero spin space 
can be obtained by taking recourse to modified Rumer-Pauling rules 
\cite{SR-VB}. Some typical VB diagrams 
are shown in Fig. \ref{figvb}.  Each orbital in the VB picture has one 
of four possibilities; (i) the orbital is empty, (ii) the orbital is 
doubly occupied, (iii) a line begins at the orbital or (iv) a line ends 
at the orbital. It is possible to associate these four possibilities of 
an orbital in a VB diagram with the four states of two bits in an integer,
identified with the orbital. Thus, for a eight orbital system, the 
VB diagrams are represented by a sixteen bit integer and these integers 
can be generated in an ascending order. In any given spin space, the 
effect of the operator term $\hat E_{ij}$ in the Hamiltonian on a basis 
state is to alter the orbital occupancy of orbitals $i$ and $j$ (subject 
to Pauli principle) and pair the spins in the orbitals that were involved 
to yield a new VB digram with a known amplitude. If the new VB diagram 
violates Rumer-Pauling rules, it is trivially possible to express it as 
a linear combination of the basis VB states.

Using the above procedure, the Hamiltonian matrix can be set-up in the
chosen total spin sector. The resulting Hamiltonian matrix is sparse
(since the number of terms in the Hamiltonian is far smaller than the
dimension of the complete VB space) and nonsymmetric and can be partially 
diagonalized to obtain a few of the low-lying eigenstates, using 
Rettrup’s modification of the Davidson algorithm, when the Hilbert 
space dimensionality is large \cite{davidson,rettrup}.
In our case, since the dimensionalities 
of the different subspaces are fairly small (the largest subspace
encountered is of dimensionality 1512 for the S=1 subspace in the A-B-A-B
system), full diagonalization of the Hamiltonian matrix is resorted to 
in each spin sector to obtain the complete eigenvalue spectrum.

\begin{figure*}[t]
\begin{center}
{\includegraphics[width=7.0cm]{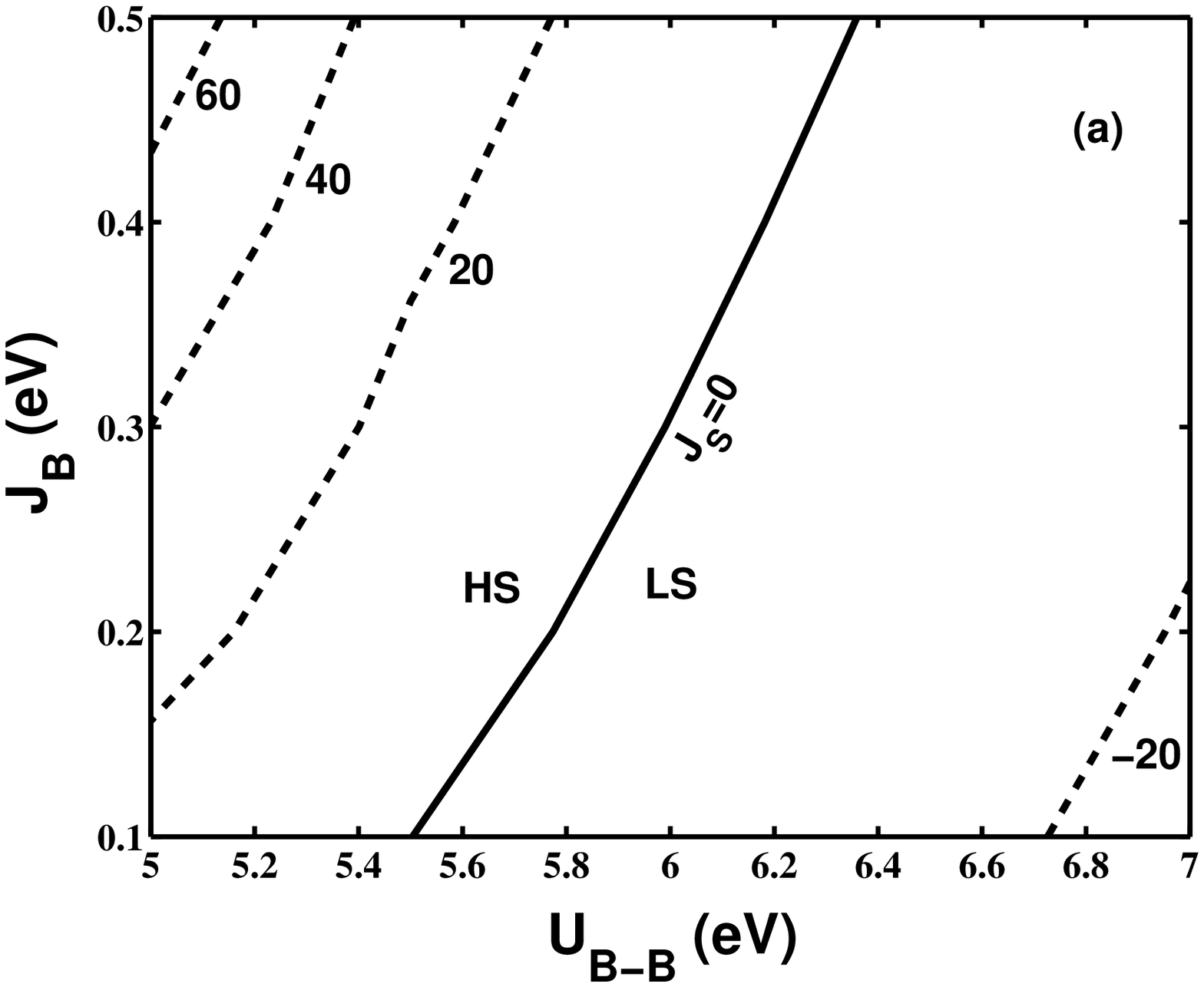}
\hspace{0.6cm} \includegraphics[width=7.0cm]{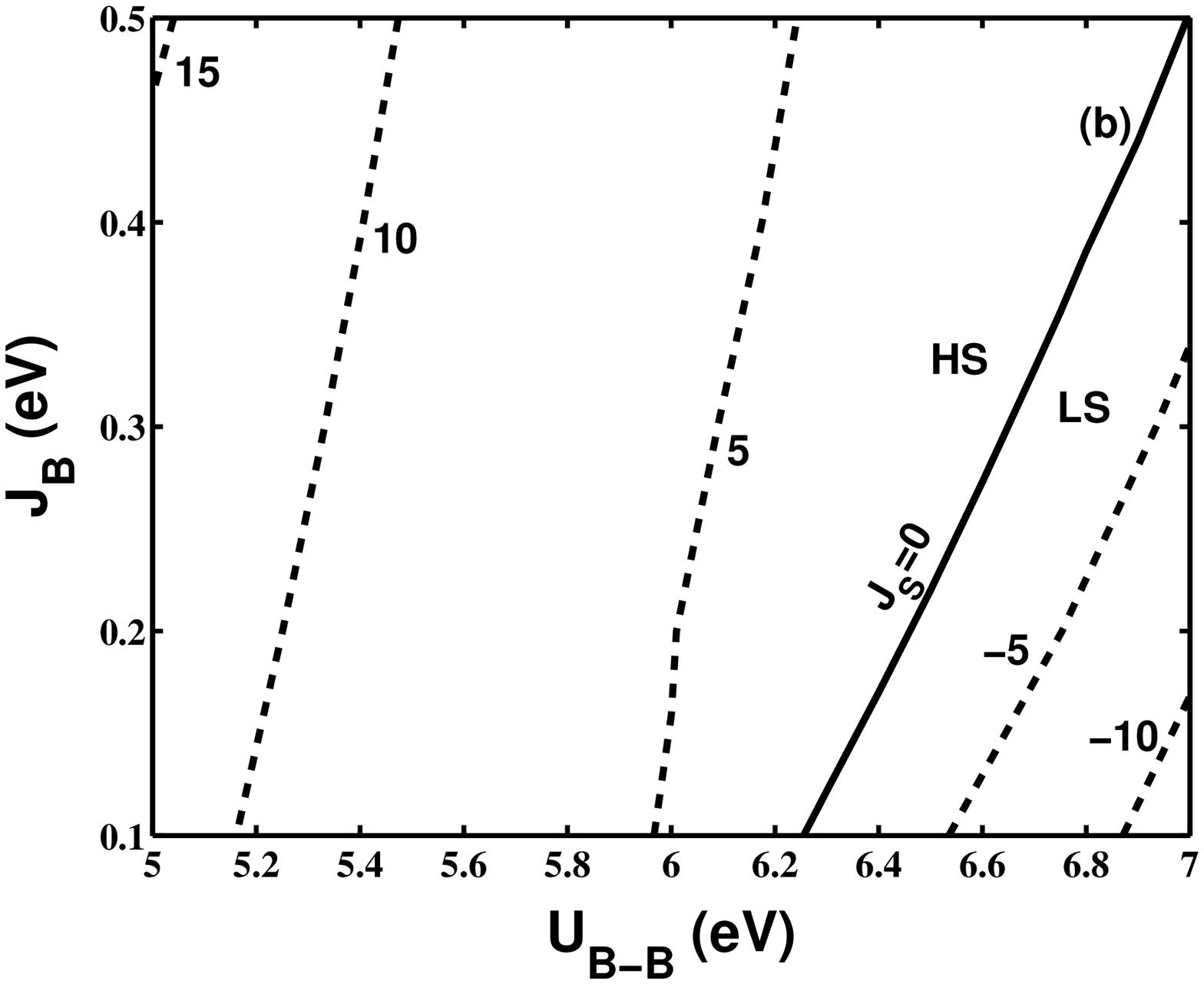}}\\
\vspace*{0.5cm}
{\includegraphics[width=7.0cm]{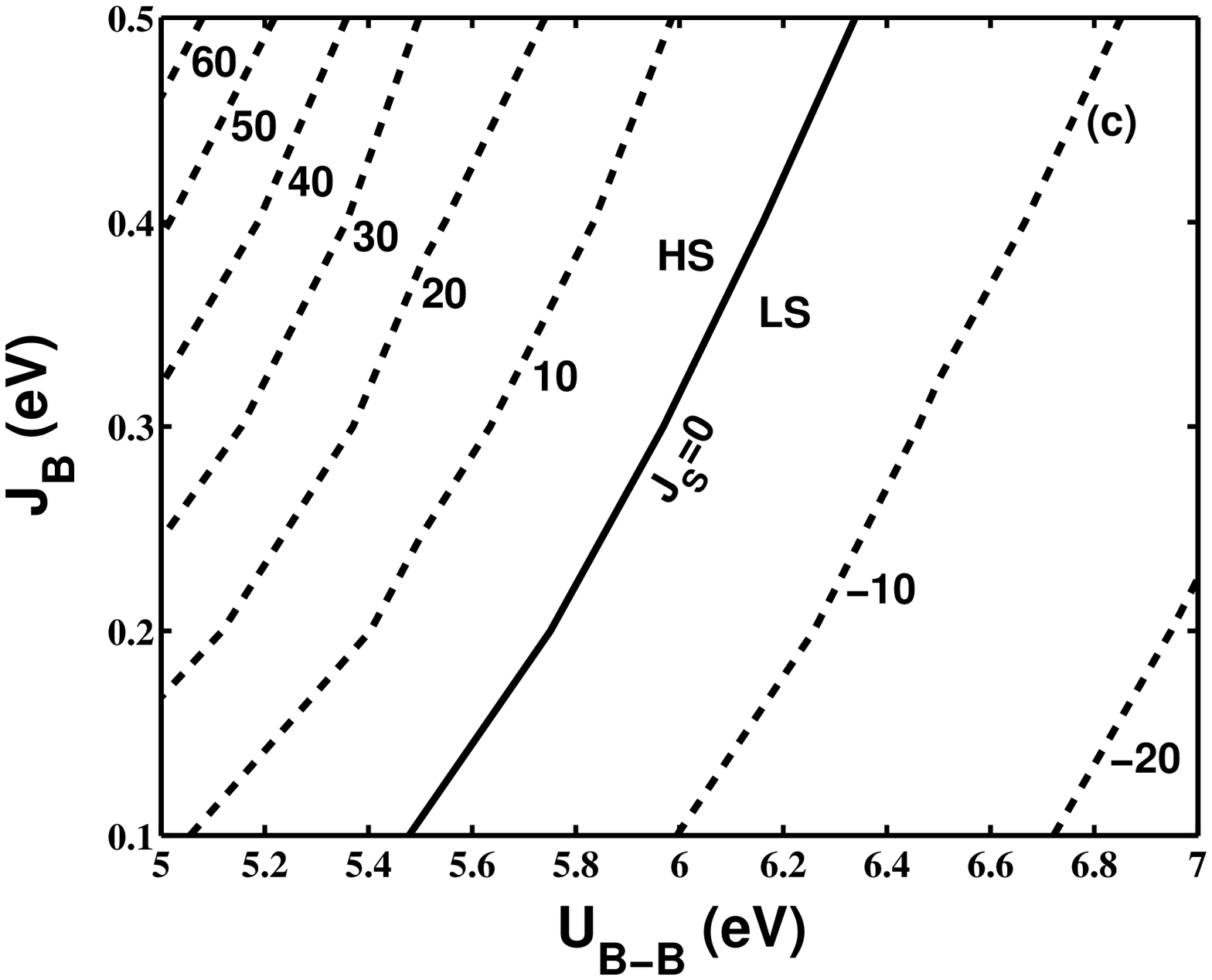}}
\hspace{0.6cm}{\includegraphics[width=7.0cm]{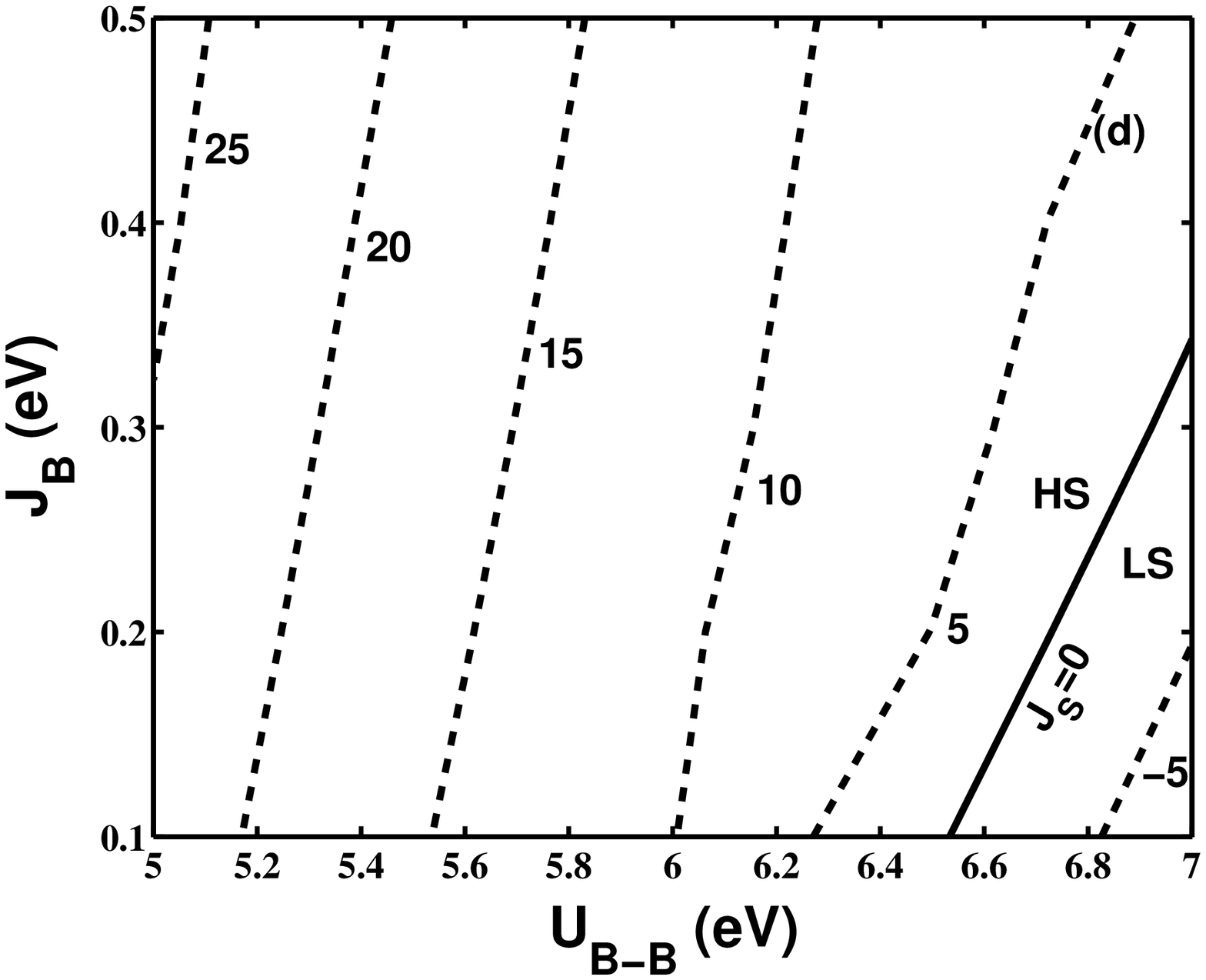}}
\end{center}
\caption[]{Contours of the effective superexchange constants ($J_S$ cm$^{-1}$) 
of (a) $A-B$, (b) $A-B-A$, (c) $B-A-B$ chains and (d) cyclic $A-B-A-B$ 
systems as a function of $U_{B-B}$ and $J_B$. The phase diagrams 
are obtained for $t$=0.1 eV, $\Delta$=0.0 eV; 
$\delta_B$=0.0 eV; $U_A$=6 eV; $U_B$=8 eV; $U_{A-A}$=4 eV; $J_A$=0.7 eV; 
$W_A$=$W_B$=1 eV. All the systems display high-spin ground state at 
higher $J_B$ and lower $U_{B-B}$ values.}
{\label{3d-UBB-JB}}
\end{figure*}

\begin{figure*}[t]
\begin{center}
{\includegraphics[width=7.0cm]{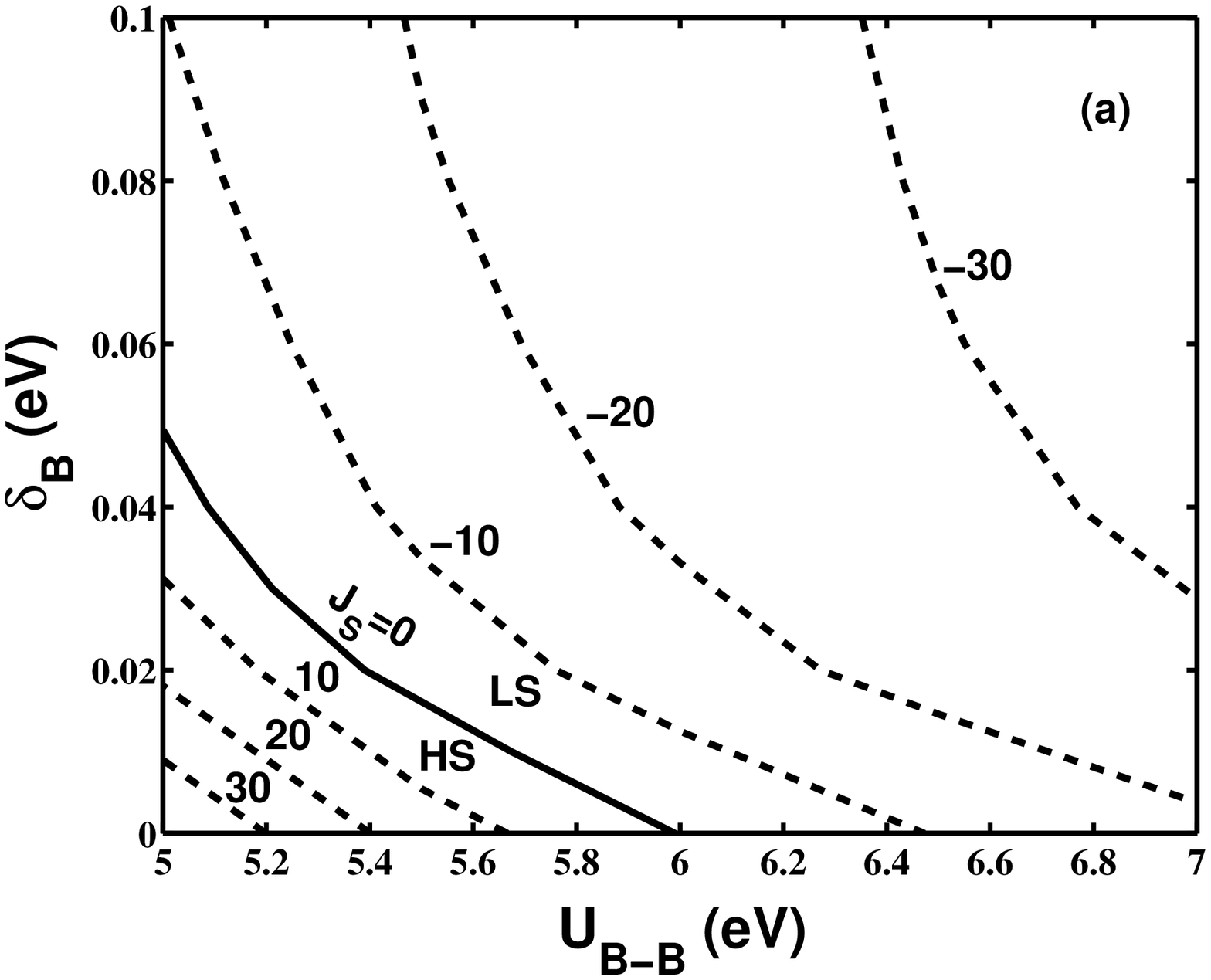}
\hspace{0.6cm} \includegraphics[width=7.0cm]{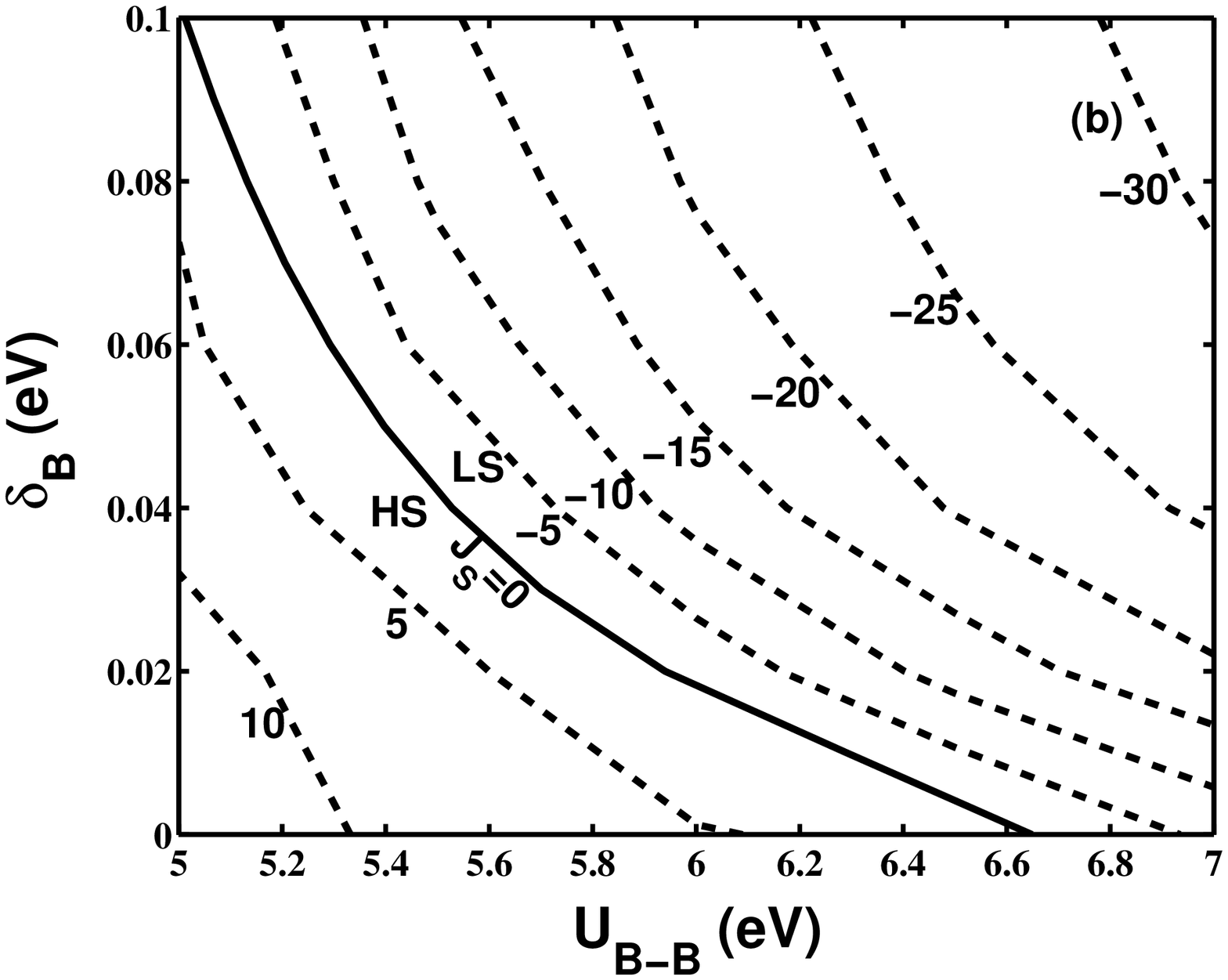}}\\
\vspace*{0.5cm}
{\includegraphics[width=7.0cm]{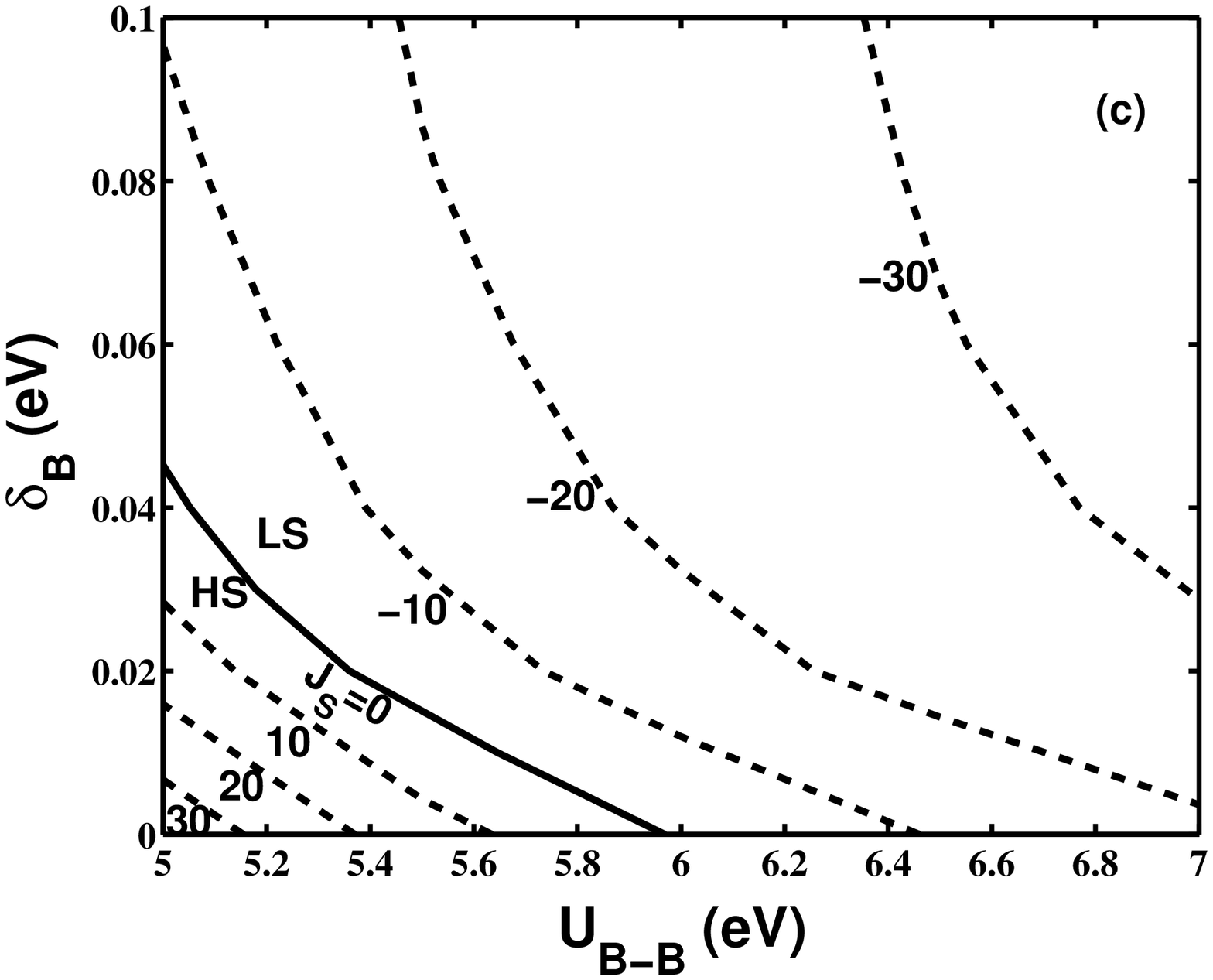}}
\hspace{0.6cm}{\includegraphics[width=7.0cm]{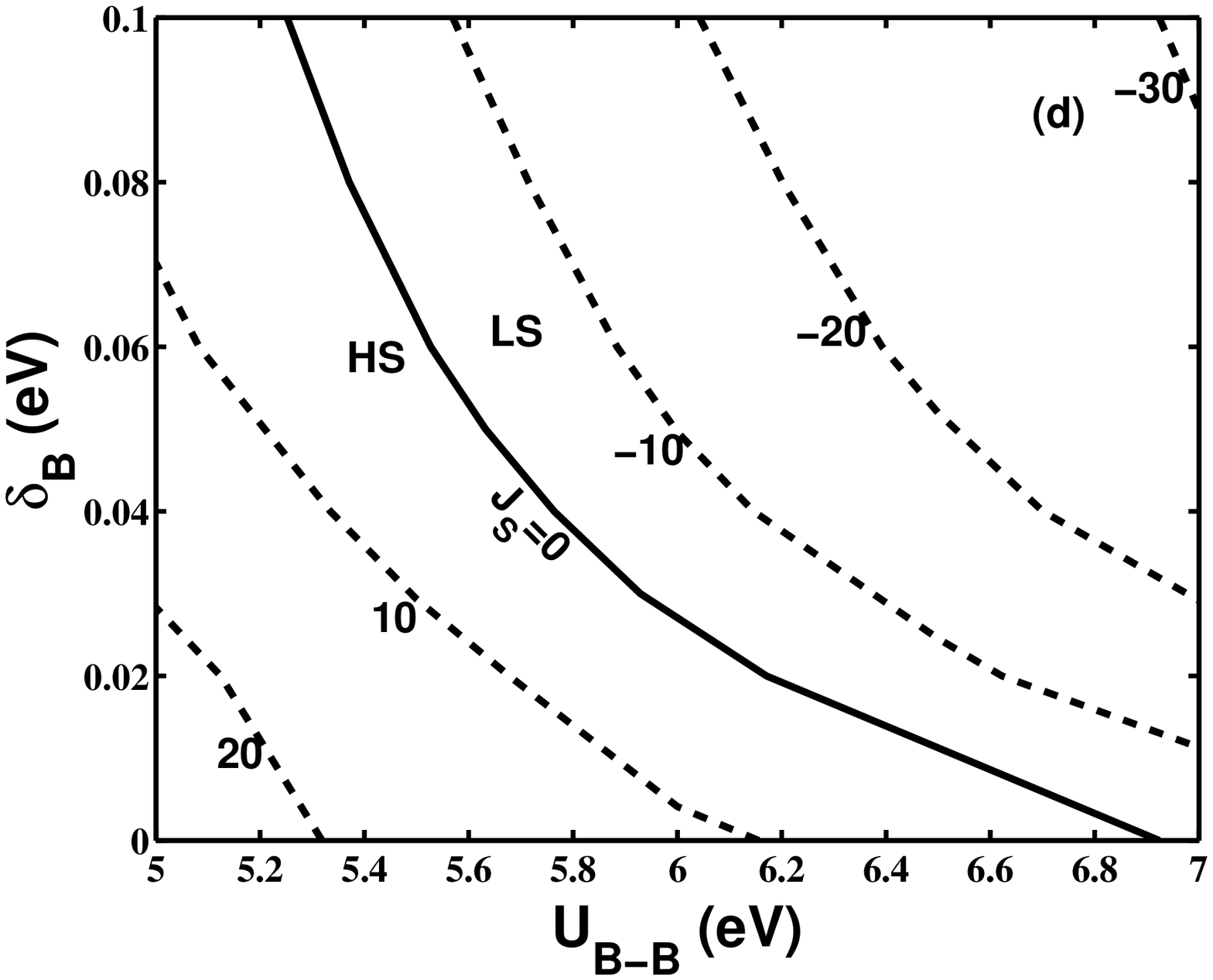}}
\end{center}
\caption[]{Contours of the effective superexchange constants ($J_S$ cm$^{-1}$) 
of (a) $A-B$, (b) $A-B-A$, (c) $B-A-B$ chains and (d) cyclic $A-B-A-B$ systems 
as a function of $\delta_B$ and $U_{B-B}$. The phase diagrams
are obtained for $t$=0.1 eV, $\Delta$=0.0 eV; $U_A$=6 eV; $U_B$=8 eV;
$U_{A-A}$=4 eV; $J_A$=0.7 eV; $J_B$=0.3 eV; $W_A$=$W_B$=1 eV. All the
systems display high-spin ground state at lower $\delta_B$ and $U_{B-B}$
values.}
{\label{3d-UBB-deltaB}}
\end{figure*}

\begin{figure*}[t]
\begin{center}
{\includegraphics[width=7.0cm]{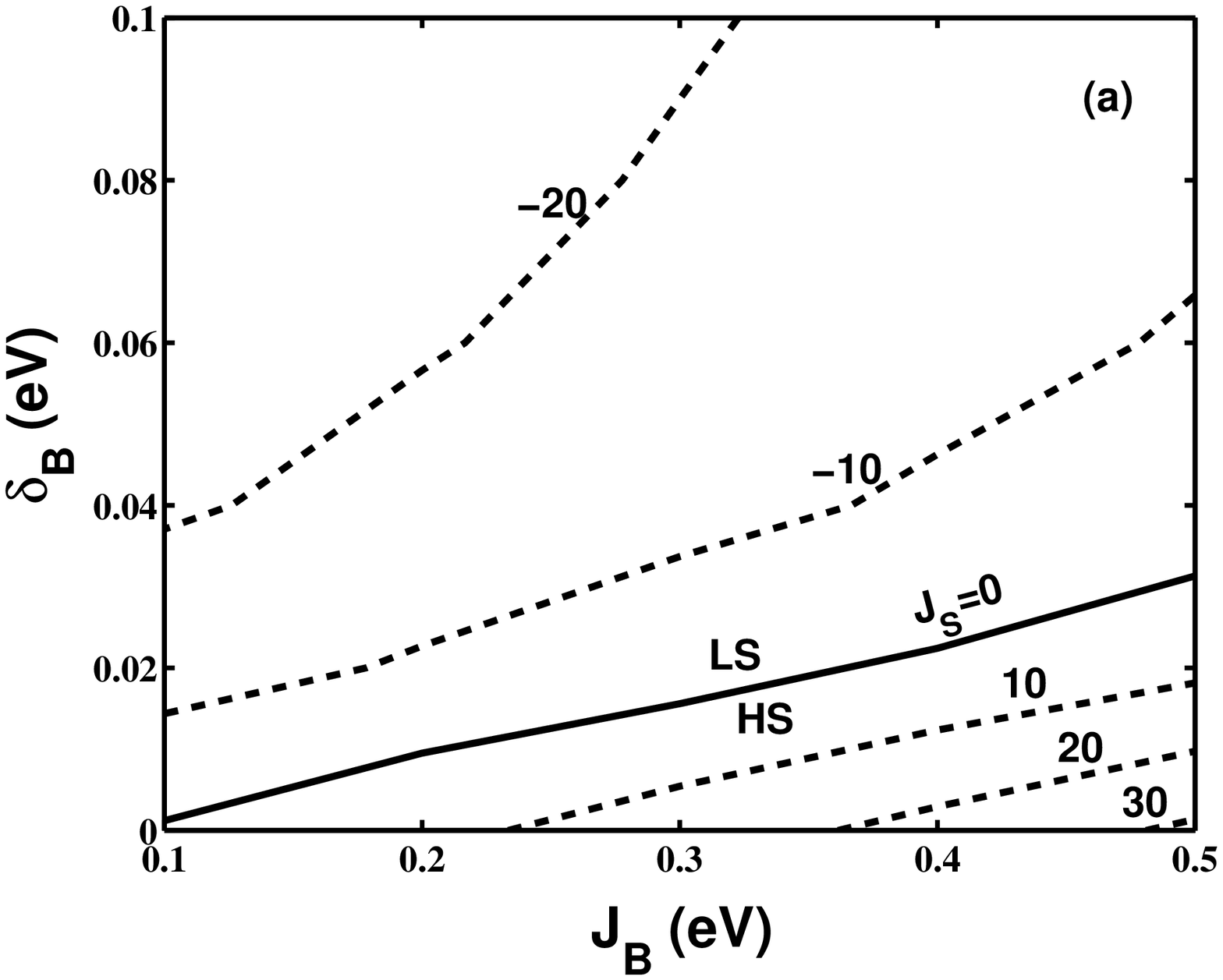}
\hspace{0.6cm} \includegraphics[width=7.0cm]{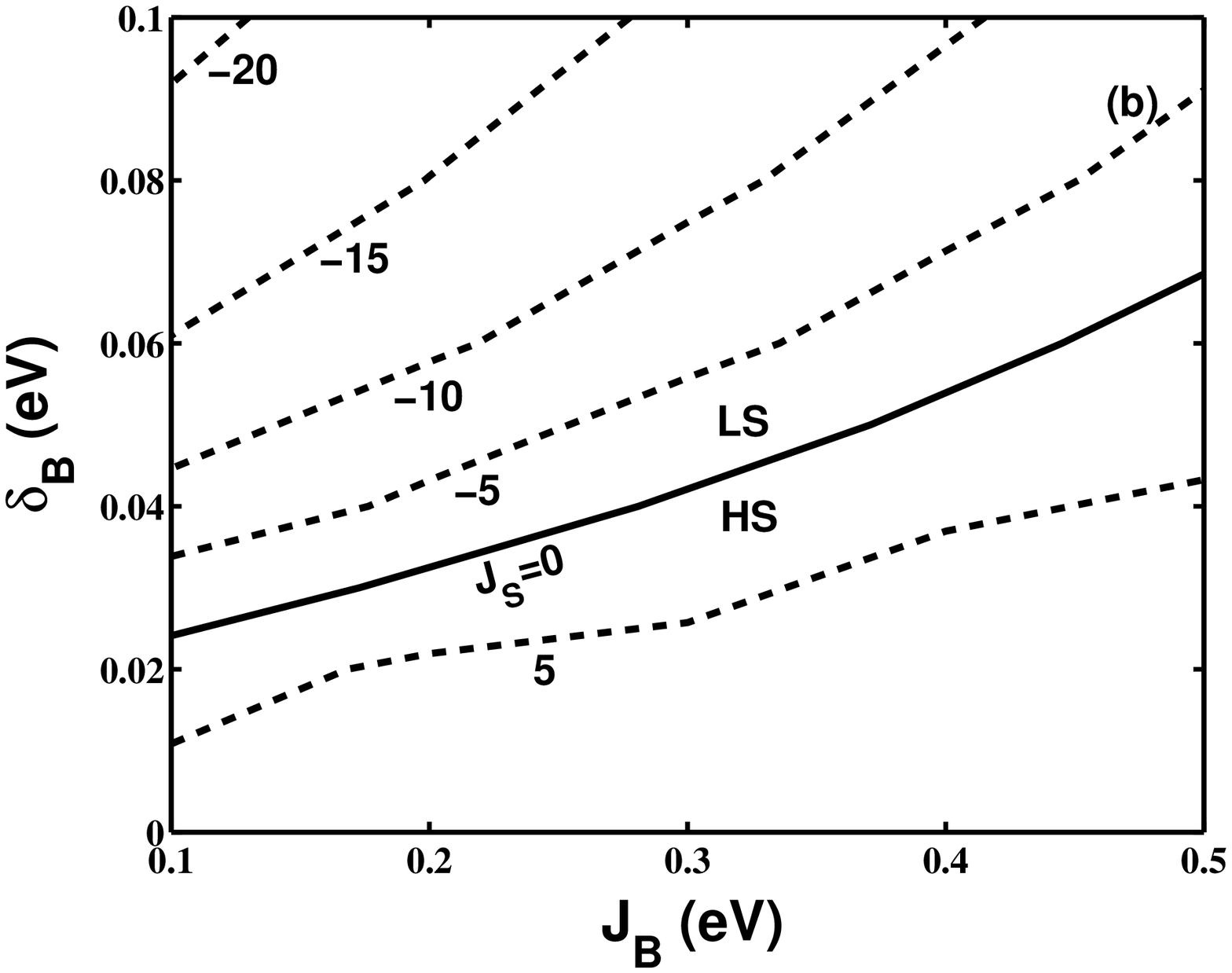}}\\
\vspace*{0.5cm}
{\includegraphics[width=7.0cm]{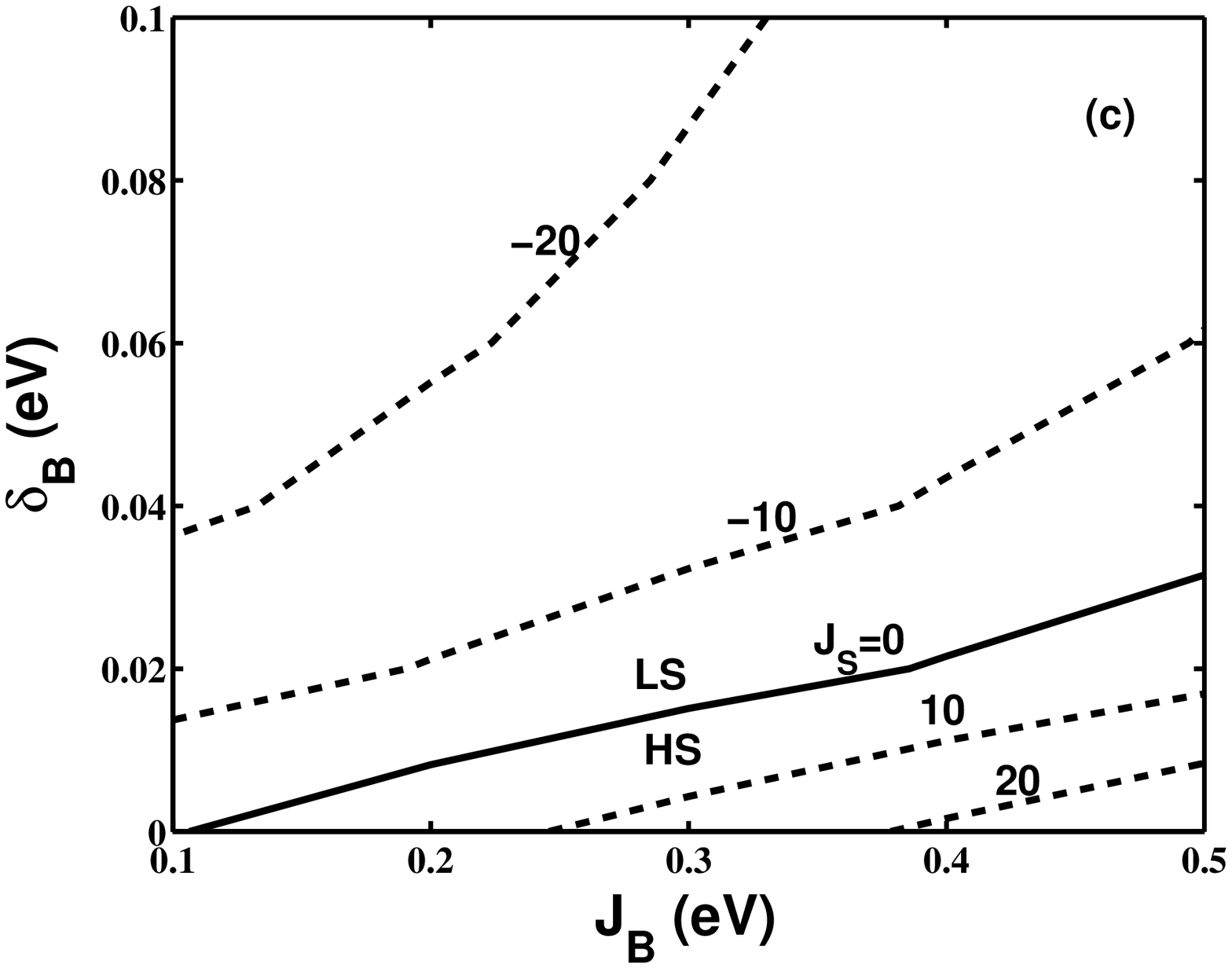}}
\hspace{0.6cm}{\includegraphics[width=7.0cm]{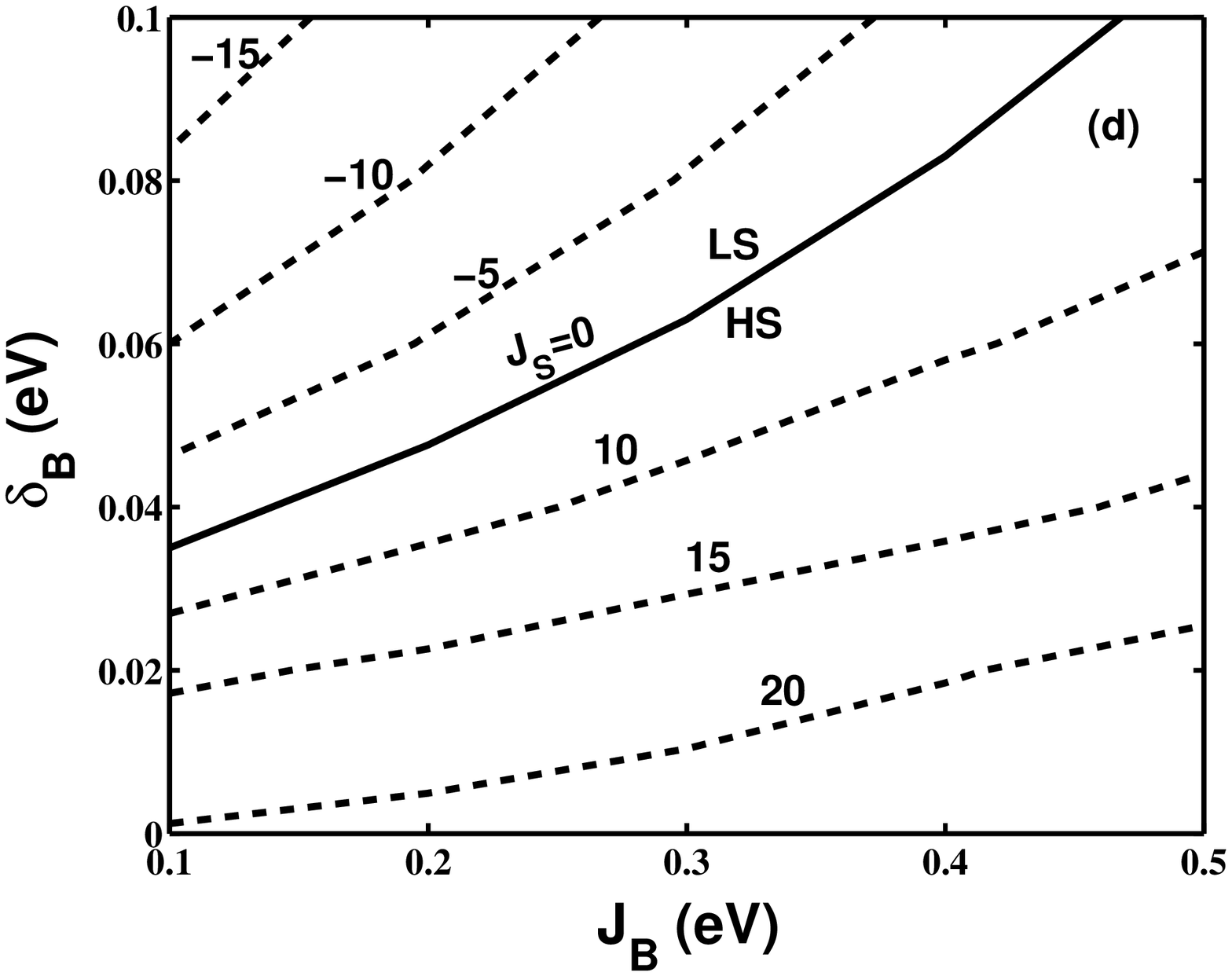}}
\end{center}
\caption[]{Contours of the effective superexchange constants ($J_S$ cm$^{-1}$) 
of (a) $A-B$, (b) $A-B-A$, (c) $B-A-B$ chains and (d) cyclic $A-B-A-B$ systems 
as a function of $J_B$ and $\delta_B$. The phase diagrams are obtained for 
$t$=0.1 eV, $\Delta$=0.0 eV; $U_A$=6 eV; $U_B$=8 eV; $U_{A-A}$=4 eV; 
$U_{B-B}$=5.5 eV; $J_A$=0.7 eV; $W_A$=$W_B$=1 eV. All the
systems display high-spin ground state at lower $\delta_B$ and higher $J_B$
values. }
{\label{3d-deltaB-JB}}
\end{figure*}

\section{IV. Results and Discussion}
The range of values on-site electron repulsion integrals take is  known 
for the transition metal ions and it is reasonable to assume that $U_A$, 
when $A$ represents Ni$^{2+}$ is smaller than $U_B$ when $B$ corresponds 
to M $(Mo^V,~W^V,~Nb^{IV})$, since the higher charge on the M ion should 
make the orbitals more compact, resulting in larger intra-orbital electron 
repulsion integrals. It is reasonable to assume $U_A$ to be 6 eV and 
$U_B$ to be 8 eV. Indeed, we have also verified that the results we 
present do not change qualitatively when these parameters are allowed 
to vary by up to 2 eV about the mean values. The parameters $U_{A-A}$ and 
$U_{B-B}$ are slightly smaller than the corresponding intra-orbital repulsion
integral. The integrals $W_A$ and $W_B$ are much smaller than the
$U_{A-A}$ or the $U_{B-B}$ integrals as they involve overlap charge 
densities and both $W_A$ and $W_B$ have been fixed at 1 eV. 
The exchange integrals $J_A$ 
and $J_B$ are somewhat smaller than the integrals $W_A$ and $W_B$. We 
also note that since it is experimentally known that the spin on the $A$ 
ion ($Ni^{2+}$) is always one we fix $J_A$ at a somewhat large value of
0.7 eV. Large $J_A$ 
reduces the repulsion between electrons with parallel spin alignment 
compared to antiparallel spin alignment in the two orbitals on 
site $A$ leading to a spin-1 object on site $A$ and this holds true
even when the degeneracy of the two orbitals on the $A$ site is slightly
lifted. For this reason, we have assumed the two orbitals on $A$ to be
degenerate. In each case, we 
have also verified that the expectation value of $\hat S_A^2$ operator
is nearly 2.0 confirming that the spin on the $A$ site is very nearly
one in all cases.  The total spin of the system in the ground state is 
sensitive to the 
parameters $\delta_B$, $J_B$ and $U_{B-B}$. A large $\delta_B$ would 
result in an antiferromagnetic exchange interaction since the virtual
state with a doubly occupied lower orbital on the $B$ site has a lower 
energy than spin one state on the $B$ ion. A large $J_B$ would however
lower the energy of the virtual state in which the $B$ ion has a spin one
configuration. A small $U_{B-B}$ would also favour a spin one virtual 
state on the $B$ ion by favouring single occupancy of the two active
$B$ ion orbitals. We have solved the model Hamiltonian over a wide range 
of the parameters and have obtained the quantum phase diagrams for 
demarcating the low-spin and high-spin ground states in this parameter space.

In Figures \ref{3d-UBB-JB}, \ref{3d-UBB-deltaB} and \ref{3d-deltaB-JB} 
we have presented the contours of superexchange values ($J_S$) in these 
systems for various values of the model parameters obtained by fitting the 
low-spin - high-spin gap to a spin Heisenberg Hamiltonian involving spin-1's 
at $A$ sites and spin-1/2's at $B$ sites. The solid line
corresponding to $J_S~=~0$ provides the phase boundary between the
high-spin and low-spin ground states. The contours corresponding to 
fixed $J_S$ values and represented by dotted lines are obtained by 
spline interpolation using MATLAB. We note that small $U_{B-B}$,
small $\delta_B$ and large $J_B$ values promote a high spin ground state while
large $U_{B-B}$, large $\delta_B$ and small $J_B$ values promote a low spin 
ground state. We also note that the superexchange $J_S$ values, for the same
model parameters, are larger for smaller system sizes.

\begin{figure*}[t]
\begin{center}
{\includegraphics[width=6.8cm]{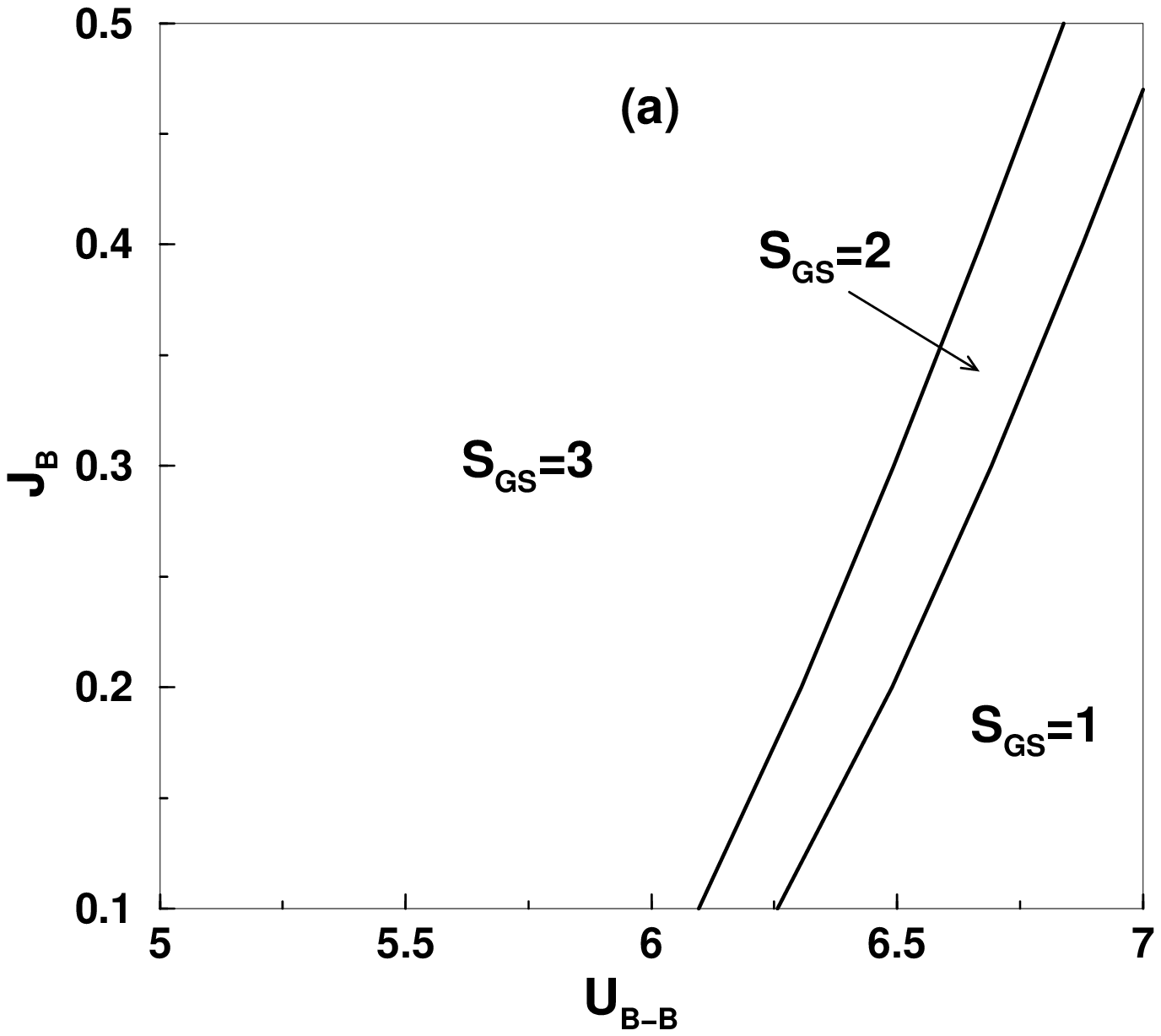}
\hspace{0.6cm} \includegraphics[width=7.0cm]{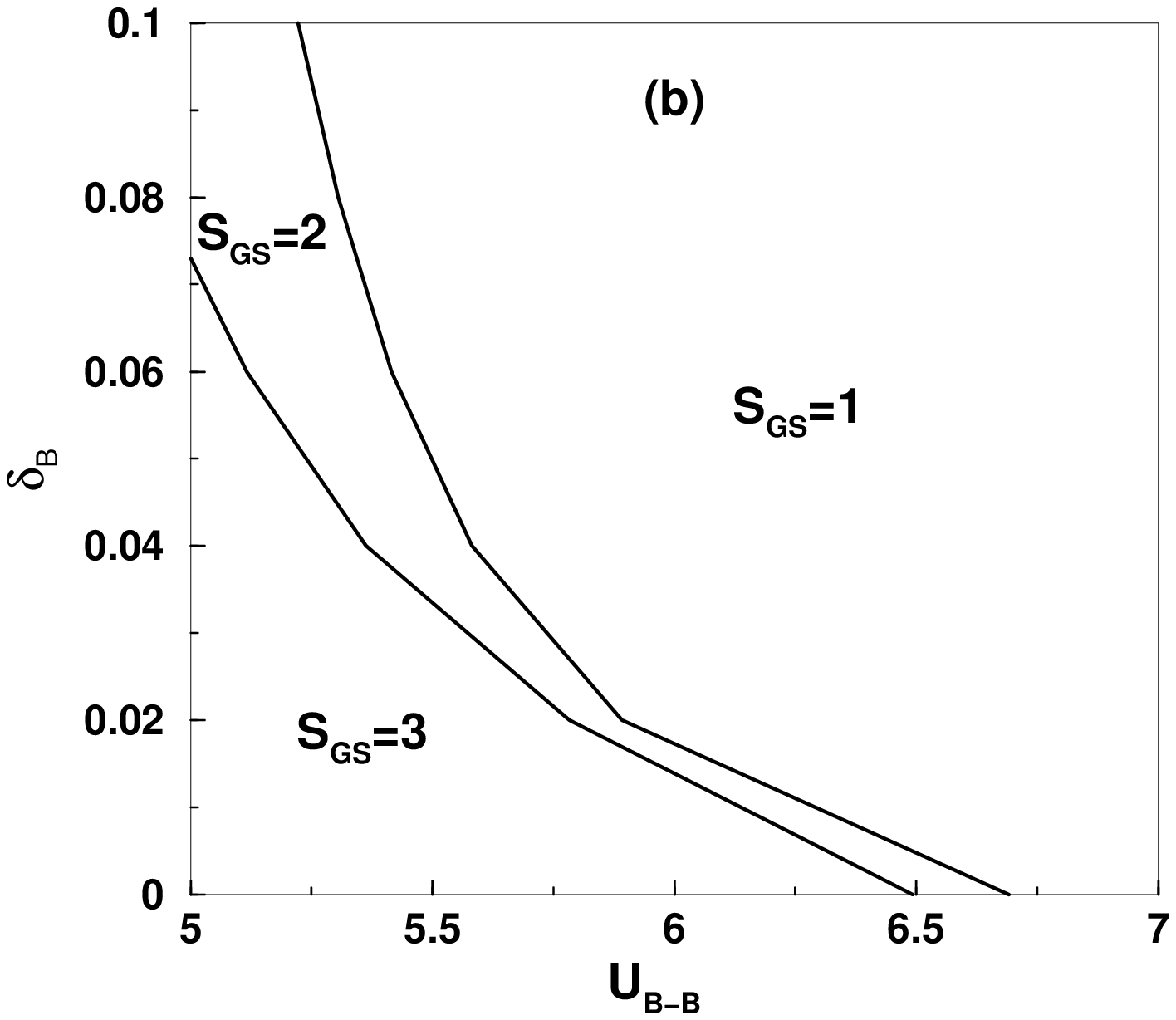}}\\
\vspace*{0.5cm}
{\includegraphics[width=7.0cm]{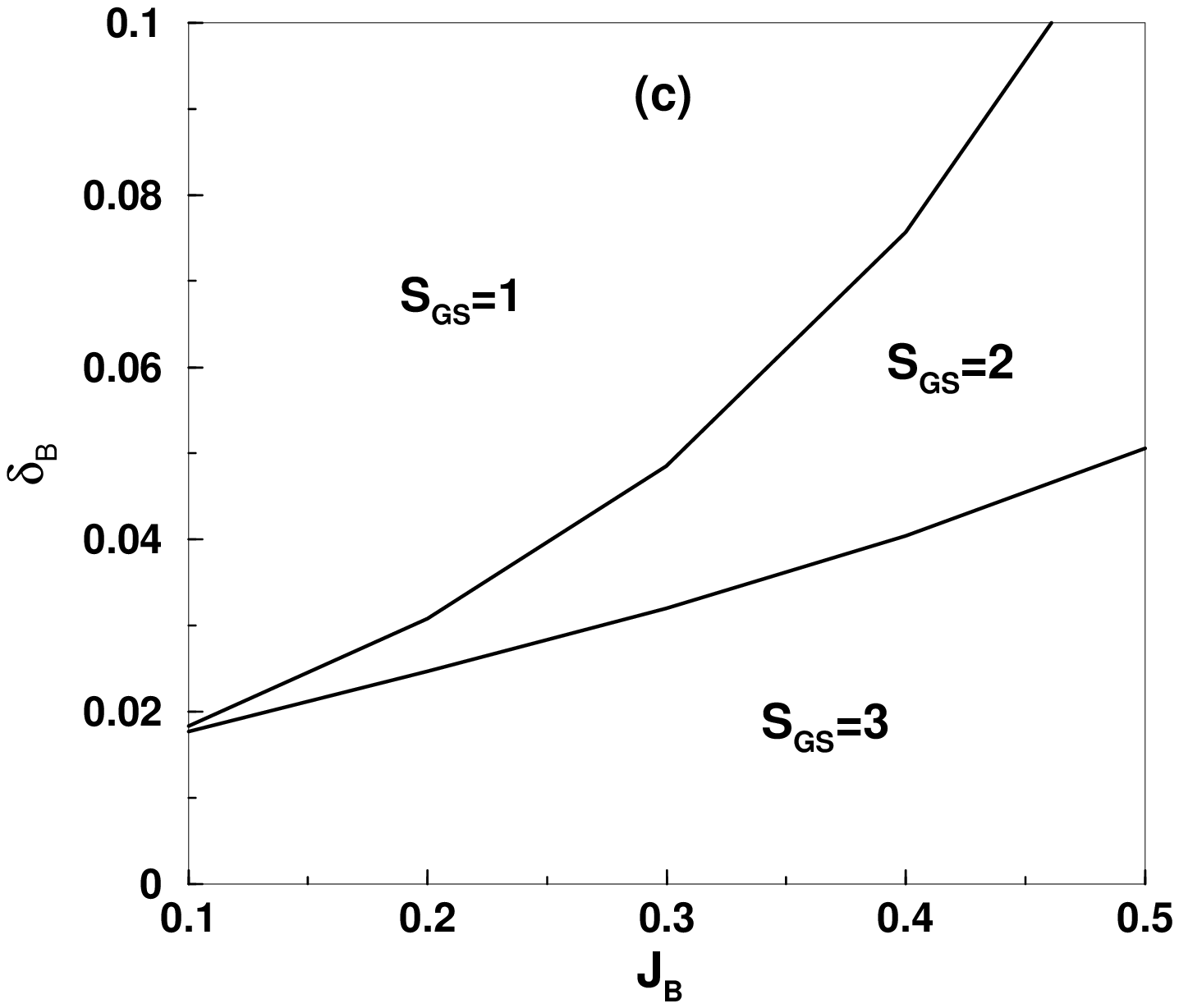}}
\end{center}
\caption[]{Phase diagrams of linear $A-B-A-B$ system in (a) $U_{B-B}~-~J_B$ 
(b) $U_{B-B}~-~\delta_B$ and (c) $J_B~-~\delta_B$ planes. $S_{GS}$ is the 
ground state spin of the system. It is interesting 
to see the appearance of intermediate spin ground state $S_{GS}=2$ for a wide 
range of parameter values, which is not prominent in the cyclic $A-B-A-B$ system.}
{\label{linearABAB}}
\end{figure*}

\subsection{A. Phase diagram in the $U_{B-B}~-~J_B$ plane}

In the $U_{B-B}~-~J_B$ plane, we should expect to see the high-spin
ground state for large $J_B$ and small $U_{B-B}$. This is because the
large exchange integral on the B site will favour parallel alignment 
of the electrons on the $B$ site, if the virtual transfer of the 
electron does not lead to a doubly occupied $B$ site orbital. The 
latter requirement is guaranteed if $U_{B-B}$ is small. In Fig. 
\ref{3d-UBB-JB} we see that in all the cases, the ground state 
corresponds to the high spin state for small $U_{B-B}$ and large 
$J_B$. We also note from the shift in the phase line to the right,
that with increase in system size the high-spin state is favoured 
for larger $U_{B-B}$ values. From the gap between the high spin and
the low spin states, we have also calculated the effective superexchange
parameter. We find that the largest ferromagnetic superexchange $J$
value is 101.42 $K$ ($70.49~cm^{-1}$) for $U_{B-B}=5$ eV and $J_B=0.5$ 
eV, while the largest antiferromagnetic superexchange $J$ value corresponds 
to -33.36 $K$ ($-23.19~cm^{-1}$) for $U_{B-B}=7$ eV and $J_B=0.1$ eV in 
the linear $A-B$ cluster. Another interesting feature to note is 
that in the $B-A-B$ cluster, the low-spin ground state appears for a much 
smaller $U_{B-B}$ value, for a given $J_B$ than in the $A-B-A$ cluster.

\subsection{B. Phase diagram in the $U_{B-B}~-~\delta_B$ plane}
The phase diagram in $U_{B-B}~-~\delta_B$ plane for all the four systems 
is shown in Fig. \ref{3d-UBB-deltaB}. We note that the ground state spin
is extremely sensitive to the $\delta_B$ value. Even a small splitting
of the $B$ site orbitals forces the system into a low-spin ground state.
This is because the lifting of the degeneracy of the $B$ site orbital
favours doubly occupied lower energy orbital in the virtual state,
which would result in stabilization of the low-spin state. Here 
again, we note that the high spin ground state is more robust in 
the larger clusters. In the cyclic $A-B-A-B$ cluster, for $U_{B-B}$ 
= 5.5 eV, the ground state shifts to low-spin state for $\delta_B$ = 0.063
eV while in the $A-B-A$ cluster, this occurs at $\delta_B$ = 0.04 eV
and in the $A-B$ system the ground state ceases to be the high spin
state for $\delta_B ~ > ~ 0.016$ eV. The highest ferromagnetic superexchange
constant $J$ is observed for $\delta_B = 0$ and $U_{B-B}$ = 5 eV while
the highest antiferromagnetic $J$ is observed for $\delta_B = 0.1$ eV
and $U_{B-B}$ = 7 eV in all the clusters. We also can note that for a given
$\delta_B$ value, we have a low spin ground state in the $B-A-B$ system, 
at much smaller value of $U_{B-B}$ value than in the $A-B-A$ system.

\subsection{C. Phase diagram in the $J_B~-~\delta_B$ plane}

Large values of $\delta_B$ has the effect of promoting the antiferromagnetic
superexchange while a large $J_B$ favours ferromagnetic superexchange. Thus, 
we see from Fig. \ref{3d-deltaB-JB} that the high spin state is the ground state
below the phase line while the low-spin state is the ground state above the
phase line. The phase line shifts higher in the $J_B~-~\delta_B$ plane as the 
system size is increased thereby showing that the high spin ground state is 
more stable to lifting of the degeneracy of the orbitals on the $B$ site for 
larger system sizes. In this plane, the behaviour of the 
$A-B-A$ and the $B-A-B$ systems are almost identical.

One very interesting feature not clearly seen in the phase diagrams of the 
cyclic $A-B-A-B$ cluster is the appearance of the intermediate spin S=2 
ground state over a very narrow range of parameter
values between the S=1 and the S=3 ground states. The intermediate 
spin ground state is seen only in the largest cluster we have studied. 
In fact, in the linear $A-B-A-B$ cluster, this region extends over a wider 
parameter values, as seen in Fig \ref{linearABAB}. 
Thus, it may be possible to synthesize high nuclearity complexes in 
intermediate spin ground states. The physics behind the existence of 
such ground states is however somewhat different from the frustrated 
magnetic exchanges present in SMMs. In our case, it is difficult to 
identify the exchange interaction between two sites as either 
ferro or antiferromagnetic and magnetism can be viewed as a whole and 
not pair-wise, as is usually the case. Besides, it is also not possible 
to map our model onto a simple Heisenberg exchange Hamiltonian.

To conclude, we have developed a model Hamiltonian which admits both low-spin
and high-spin ground states for small changes in the values of the model
parameters. The only model parameter that can perhaps be determined directly 
for a system is $\delta_B$, from spectroscopic data. Other parameters can only 
be inferred indirectly either from electron spectroscopic studies on 
simpler systems or from abinitio model system calculations. 
Our model can explain the observed antiferromagnetic exchange
in the Nb$_6$Ni$_{12}$ and related systems, contrary to superexchange
rules \cite{NbNi,sutter-unp}. The model also yields reasonable effective 
superexchange constants for model parameters in the accepted range.

\section{Acknowledgment}

The authors thank DST for the support received under a joint Indo-French
Laboratory for Solid State Chemistry (IFLaSC) and Indo-French Centre for
Promotion of Advanced Research (IFCPAR)/Centre Franco-Indien pour la 
Promotion de la Recherche Avanc\'ee (CEFIPRA) for generous support under
Project 3108-3 on Design, Synthesis and  Modeling Molecular Magnets.


\begin{thebibliography}{99}
\bibitem{footnote} The splitting of the degenerate orbitals on $A$ 
is expected to be small as the $A$ ion is always in spin-1 state and the direct 
exchange integral involving these orbitals on $A$ is large enough to ensure that 
such splitting does not result in spin pairing on the $A$ site.
\end{thebibliography}
\end{document}